\documentclass[preprint,12pt]{aastex}
\usepackage{amsmath} \usepackage{amsfonts} \usepackage{amssymb}  \usepackage{apjfonts}\usepackage{color}

\pdfoutput = 1

\shorttitle{Escape of heavy atoms}
\shortauthors{T.T. Koskinen et al.}

\begin{document}

\title{Thermal escape from extrasolar giant planets}

\author{T. T. Koskinen\altaffilmark{1}, P. Lavvas\altaffilmark{3}, M. J. Harris\altaffilmark{2}, R. V. Yelle\altaffilmark{1}}
\altaffiltext{1}{Lunar and Planetary Laboratory, University of Arizona, 1629 E. University Blvd., Tucson, AZ 85721--0092; tommi@lpl.arizona.edu}
\altaffiltext{2}{Department of Physics and Astronomy, University College London, Gower Street, London WC1E 6BT, UK}
\altaffiltext{3}{Groupe de Spectrom\'etrie Mol\'eculaire et Atmosph\'erique UMR CNRS 6089, Universit\'e Reims Champagne-Ardenne, 51687 Reims, France}

\begin{abstract}   
The detection of hot atomic hydrogen and heavy atoms and ions at high altitudes around close-in extrasolar giant planets (EGPs) such as HD209458b imply that these planets have hot and rapidly escaping atmospheres that extend to several planetary radii.  These characteristics, however, cannot be generalized to all close-in EGPs.  The thermal escape mechanism and mass loss rate from EGPs depend on a complex interplay between photochemistry and radiative transfer driven by the stellar UV radiation.  In this work we explore how these processes change under different levels of irradiation on giant planets with different characteristics.  We confirm that there are two distinct regimes of thermal escape from EGPs, and that the transition between these regimes is relatively sharp.  Our results have implications on thermal mass loss rates from different EGPs that we discuss in the context of currently known planets and the detectability of their upper atmospheres.                        
\end{abstract}


\section{Introduction}
\label{sc:intro}       

The upper atmospheres of three different EGPs HD209458b, HD189733b, and WASP-12b were recently probed by UV transit observations obtained by the Hubble Space Telescope (HST) \citep{vidalmadjar03,vidalmadjar04,linsky10,fossati10,lecavelier10,lecavelier12}.  These observations indicate that close-in EGPs such as HD209458b are surrounded by a hot thermosphere composed of atomic hydrogen that extends to several planetary radii and provides the required line of sight (LOS) optical depth in atomic resonance lines to explain large transit depths \citep{koskinen10,koskinen12a,koskinen12b}.  It is also widely believed that the atmospheres of close-in EGPs undergo hydrodynamic escape \citep[e.g.,][]{lammer03,yelle04}.  

The term hydrodynamic escape is commonly associated with transonic outflow \citep{parker58}.  However, recent calculations by \citet{koskinen12a} demonstrate that the altitude of the sonic point is strongly dependent on radiative transfer and heating rates in the upper atmosphere.  Furthermore, they show that on planets such as HD209458b the temperature gradient in the upper atmosphere is close to adiabatic and thus the sonic point is likely to be located at a relatively high altitude.  On the other hand, the altitude of the sonic point may also depend on poorly known factors such as tidal forces, interaction with the stellar wind, and the strength of the possible planetary magnetic field \citep[e.g.,][]{erkaev07,stone09,trammell11}.  This means that the location of the sonic point in models of thermal escape is mostly of theoretical interest.  The currently available observations do not have adequate signal to noise (S/N) or wavelength resolution to allow for the location of the sonic point to be determined on actual planets.  What is clear, however, is that rapid escape on close-in EGPs affects the temperature and density profiles in their atmospheres, and that escape from their atmospheres can be roughly modeled with fluid equations.  In line with \citet{tian08a,tian08b}, we consider this as the hydrodynamic escape regime.  

Needless to say, the thermospheres of the giant planets in the solar system differ significantly from this picture.  For instance, the exobase on Jupiter is located only $\sim$2,300 km above the 1 bar level \citep{strobel02}, while on Saturn it is located only 2700--3000 km above the 1 bar level \citep{koskinen13}.  In both cases the thermal escape parameter $X = G M_p m/k T r_{\text{exo}}$ is large, between 250--450, and thermal escape is firmly in the classical Jeans regime based on kinetic theory \citep{volkov11a,volkov11b}.  In this case escape does not affect the temperature and density structure of the atmosphere significantly, and proceeds on particle by particle basis from the exobase into the nearly collisionless exosphere.  Given this background,  it is interesting to assess how the escape mechanism and thermal escape rates on different EGPs orbiting Sun-like stars vary with orbital distances.  The upper atmospheres of close-in giant planets are heated by the stellar UV radiation, and at some orbital distance the flux should be high enough to cause a transition from kinetic escape to hydrodynamic escape.

\citet{koskinen07a,koskinen07b} used results from a three-dimensional model for the thermospheres of EGPs to argue that this transition, which is controlled by the dissociation of H$_2$ below the exobase and the subsequent lack of efficient infrared cooling by H$_3^+$, occurs within a surprisingly narrow range of orbital distances between 0.1 and 0.2 AU for Jupiter-type planets orbiting Sun-like stars.  In other words there is a relatively sharp limit to the incoming UV flux that causes the atmosphere to expand and begin to undergo rapid escape.  These calculations also indicate that the upper atmospheres of close-in EGPs are quite resilient -- they remain relatively cool and stable at surprisingly small orbital distances.  The results provide an important perspective on the predicted properties of the several hundreds of known EGPs, and can be used to guide statistical studies of mass loss from EGPs \citep[e.g.,][]{lecavelier07,lammer09} as well as to search for good targets for future observations of their upper atmospheres. 

Previous work by \citet{koskinen07a,koskinen07b} relied on a model composed exclusively of hydrogen and helium.  However, recent calculations have shown that the photochemistry of carbon and oxygen species may play an important role in controlling the composition in the upper atmosphere of HD209458b \citep{moses11}.  In particular, complex photochemistry in a solar composition atmosphere can lead to significant dissociation of H$_2$ by OH radicals at a relatively deep pressure level of 1 $\mu$bar.  It should be noted that both on Jupiter and Saturn the thermosphere is composed mostly of H$_2$ up to the exobase that is located at a pressure of a few pbar, so this is another important difference between close-in EGPs and the giant planets in the solar system.  

In this paper we update the previous calculations of \citet{koskinen07a,koskinen07b} to include more complex photochemistry.  More specifically, we study the dissociation chemistry of H$_2$ as a function of orbital distance on EGPs by relying on existing thermal structure calculations and new photochemical calculations.  We use the results to assess the escape mechanism and mass loss rates under different levels of irradiation based on new theoretical results \citep{volkov11a,volkov11b} and their application to close-in EGPs \citep{koskinen12a,koskinen12b}.  We then proceed to generalize the results to a sample of known EGPs to make specific predictions about the nature of their upper atmospheres.  

          
\section{Methods}
\label{sc:methods}     


We studied the dissociation chemistry of H$_2$ by using the photochemical model of \citet{lavvas08a,lavvas08b} that was recently modified to simulate the atmospheres of gas giant planets at high temperatures \citep[e.g.,][]{koskinen12a,koskinen12b}.  In addition to the reaction rate coefficients, the required inputs for the photochemical model are the stellar spectrum, a temperature-pressure (T-P) profile, and the eddy diffusion coefficient $K_{zz}$ that generally depends on altitude.  In the lower atmosphere we used T-P profiles from \citet{sudarsky03} and \citet{burrows04} that were calculated for EGPs at orbital distances ranging from 0.1 to 1 AU.  We calculated the composition at 1 AU, 0.5 AU, 0.4 AU, 0.3 AU, 0.2 AU, 0.1 AU, and 0.047 AU.  For HD209458b we used the T-P profile from \citet{showman09}.  The T-P profiles above the 1 $\mu$bar level of each simulation were adopted from \citet{koskinen07a,koskinen07b} for different orbital distances, and connected smoothly with the T-P profiles at lower altitudes.   We estimated the terminal values of $K_{zz}$ in the upper atmosphere based on the simple scaling proposed by \citet{koskinen10b}.  These values range from 10$^3$ to 10$^4$ m$^2$~s$^{-1}$ between 1 and 0.047 AU, respectively.  The reader should note that these values are more conservative than the relatively high values typically presented in the literature \citep[e.g.,][]{moses11}.  


We then proceeded to use the results from the photochemical model as lower boundary conditions for the escape model at the 1 $\mu$bar level, and recalculated the temperature, density, and velocity profiles in the thermosphere.  For this purpose, we used the one-dimensional escape model of \citet{koskinen12a,koskinen12b}.  We fixed the lower boundary temperature of this model to a value consistent with the T-P profiles above.  The lower boundary was placed at 1 $\mu$bar because most of the EUV radiation that powers escape in the model is absorbed above this level.  The reader should note that X-rays typically penetrate deeper than the 1 $\mu$bar level and accounting for their heating impact on the atmosphere requires a model of radiative transfer that includes cooling by the abundant molecules.  At the upper boundary we applied either the modified Jeans or Jeans conditions, depending on the value of $X$.  In both cases we included the ambipolar electric potential that can be important in ionized atmospheres \citep[e.g.,][]{garciamunoz07,koskinen12a}.  Our model is self-consistent with regard to these boundary conditions in that the altitude of the exobase is updated regularly during the simulation, and the Jeans conditions are applied slightly below the exobase where the Knudsen number $Kn \approx$~0.1.  In this way the model develops to a steady state with a converged altitude for the exobase, when an exobase exists at a reasonably low altitude.  At close-in orbits the exobase extends to very high altitudes and the modified Jeans conditions are compatible with the outflow boundary conditions \citep{koskinen12a}.  
 
\section{Results}
\label{sc:results}     

\subsection{H$_2$ dissociation chemistry}
\label{subsc:h2diss}

The photochemical calculations constrain the mixing ratio of H at the 1 $\mu$bar lower boundary of the hydrodynamic escape model.  The latter also includes the ion chemistry of H$_2$, H, and He with a reaction scheme similar to that of \citet{yelle04} and \citet{koskinen09}.  With an accurate lower boundary constraint, the escape model can therefore update the H$_2$/H transition level based on the temperature and chemistry in the thermosphere, as long as the homopause is located reasonably near the 1 $\mu$bar level.  Our calculations show that the mixing ratio of H at the 1 $\mu$bar level varies from about 4.9~$\times$~10$^{-3}$ at 1 AU to about 1.7~$\times$~10$^{-1}$ at 0.1 AU, finally reaching about 0.5 for HD209458b.  As a point of comparison, the corresponding mixing ratio at Jupiter is about 10$^{-3}$ \citep{seiff98}.  Interestingly, H$_2$ is mostly dissociated thermally at orbital distances greater than 0.1 AU. 

At close-in orbits, within 0.1 AU, H$_2$ can also be dissociated by OH radicals that are released by the photolysis of water molecules \citep[e.g.,][]{moses11}.  Because this dissociation mechanism is primarily driven by FUV radiation, that penetrates past the EUV heating peak, it can potentially dissociate H$_2$ at deeper pressure level than thermal dissociation.  We note, however, that the relative importance of thermal dissociation and photochemistry depends on the assumed T-P profile.  The dayside T-P profile typically assumed for HD209458b, which was also used in this work is warm enough to dissociate thermospheric H$_2$ anyway.  

At 1 AU the H$_2$/H transition occurs above the 0.2 nbar level i.e., above the thermospheric heating peak that is located at 1--10 nbar.  Near the exobase at 2.4 pbar, the mixing ratio of H is close to unity.  This means that there is a rather sharp H$_2$/H transition at low pressures near the exobase, but that H$_2$ is the dominant species in most of the thermosphere.  Moving inward from 1 AU to 0.4 AU, the H$_2$/H dissociation front moves down to the 3 nbar level.  By 0.3 AU, it reaches down to the 20 nbar level i.e., below the EUV heating peak.  At this point the thermosphere also heats up and expands dramatically (see Section~\ref{subsc:tempvel} below).  At 0.2 AU the exobase is above the 16 $R_{\text{p}}$ upper boundary and the maximum temperature is 10,300 K.  At 0.1 AU the H$_2$/H transition is located near the 50 nbar level and, as we pointed out above, for HD209458b this transition takes place near the 1 $\mu$bar level.   

\subsection{Temperature and velocity}
\label{subsc:tempvel}     

\begin{figure}
  \epsscale{0.8}
  \plotone{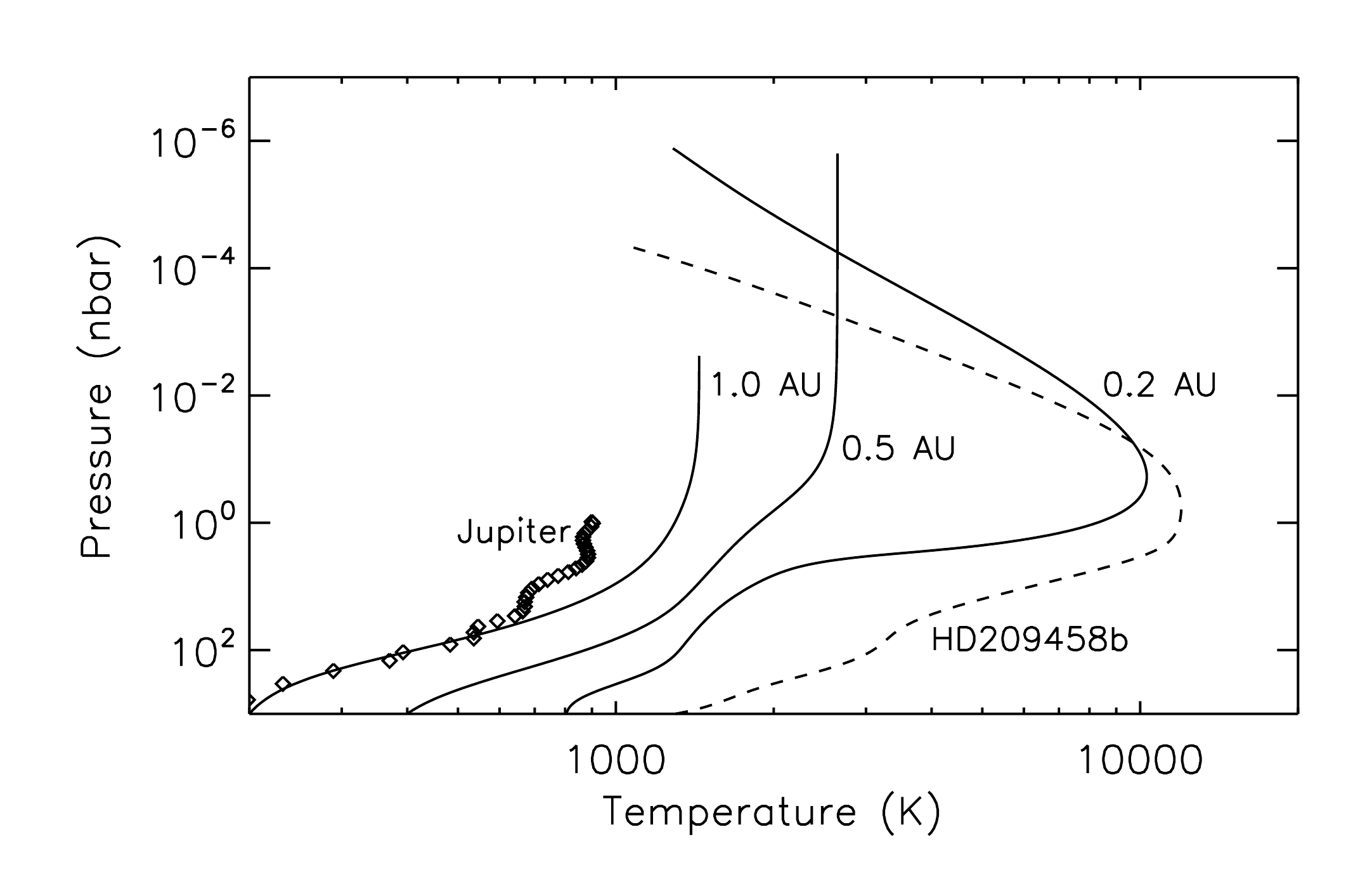}
  \caption{Simulated temperature-pressure profiles for EGPs with planetary parameters of HD209458b at different orbital distances (solid lines).  The T-P profiles are shown up to the exobase when it exists (at orbital distances greater than 0.3 AU).  Also shown are the T-P profiles for Jupiter \citep{seiff98} and the simulated profile for HD209458b \citep{koskinen12a,koskinen12b}.}
  \label{fig:tp}
\end{figure}

Figure~\ref{fig:tp} shows the T-P profiles from our new calculations for different orbital distances above the 1 $\mu$bar level.  These calculations are based on the planetary parameters of HD209458b.  The results are compared with the model T-P profile for HD209458b \citep{koskinen12a,koskinen12b} and the measured equatorial T-P profile for Jupiter \citep{seiff98}.  At 1 AU the exospheric temperature is $T_{\infty} \approx$~1440 K.  In general, the temperature increases rather steeply with altitude between 0.1 and 100 nbar in a region where most of the stellar EUV radiation is absorbed.  At 1 AU stellar heating and conduction dominate the energy balance.  Particularly at high altitudes heating is almost exactly balanced by conduction, leading to an isothermal temperature profile.  Similarly to Jupiter, radiative cooling by H$_3^+$ is also important near the heating and ionization peak.  Contrary to HD209458b, cooling by adiabatic expansion or advection are negligible.        

The exospheric temperature at 1 AU is comparable to the exospheric temperature on Earth.  The model T-P profile at 1 AU is also similar to the observed T-P profile on Jupiter.  In particular, our temperature profile agrees well with the Jovian profile below the 20 nbar level.  It is well known that the temperatures on Jupiter and other giant planets in the solar system are higher than expected from solar heating alone \citep[e.g.,][]{miller05}, and the heating mechanism responsible for this is currently unknown.  In this regard it is interesting that the correct T-P profile in the lower thermosphere can be obtained by moving Jupiter from 5 AU to about 1 AU i.e., by multiplying the solar flux by a factor of 25.    

It is not easy to predict the degree to which the atmospheres of EGPs might also be warmer than expected from solar heating.  The situation in the solar system may provide some guidance on this though.  The leading suggestions for the missing heating mechanism on Jupiter and Saturn include redistribution of auroral energy from the poles to the equator or, given the rapid rotation of these planets and polar ion drag that may constrain the auroral energy to high latitudes, direct heating by breaking gravity waves \citep[e.g.,][]{mullerwodarg06,smith07}.  Both of these phenomena are observed on Earth, but solar EUV heating is still the most important overall energy source in the thermosphere and the temperature in the Earth's thermosphere can easily be explained.  It may thus be argued that stellar heating simply overtakes any secondary heat sources within 1 AU.  We rely on this assumption in this work, and present results within 1 AU based on stellar heating only. 

Adiabiatic cooling or advection that are associated with the escape of the atmosphere do not affect the energy balance at 1 AU.  The `critical' thermal escape parameter at the exobase ($r_{\infty} =$~1.09 R$_{\text{p}}$) is $X_{\infty} =$~72, and near the EUV heating peak $X =$~190.  These values imply a negligible mass loss rate of $\dot{M} \approx$~3.7~$\times$~10$^{-23}$ kg~s$^{-1}$.  We note that the thermospheric heating efficiency at 1 AU, which is based on the balance of radiative (photoelectron) heating and cooling, is about $\eta_{\text{eff}} =$~0.48.  This value is not very different from $\eta_{\text{eff}} =$~0.44 that we estimated for HD209458b\footnote{Note that these efficiencies are defined in terms of the stellar flux at wavelengths shorter than 912 \AA.}.  Thus the energy-limited mass loss rate at 1 AU is about 10$^4$ kg~s$^{-1}$.  This means that the escape rate that we calculated at 1 AU is technically not in the energy-limited regime --  instead it is many orders of magnitude lower than the energy-limited rate.  We argue in Section~\ref{subsc:escape} that this apparent discrepancy arises from a confusion about the definition of heating efficiency rather than any new insight into the physics of evaporation.

Moving inward from 1 AU to 0.5 AU, the exospheric temperature increases from 1440 K to about 2630 K.  With this increase in temperature the exobase extends to about 1.8 R$_\text{p}$ where the pressure is about 7~$\times$~10$^{-7}$ nbar.  Such a low pressure for the exobase is possible because the thermosphere is mostly ionized at radii higher than about 1.4 R$_{\text{p}}$, and the cross section for Coulomb or ion-neutral collisions is much larger than the cross section for neutral-neutral collisions.  This is in contrast to 1 AU where the electron-neutral mixing ratio at the exobase is only $x_e =$~6~$\times$~10$^{-3}$.  As shown in Figure~\ref{fig:tp}, the temperature profile at 0.5 AU is isothermal near the exobase, indicating that heating is balanced by conduction.  Similarly to 1 AU, escape has a negligible impact on the T-P profile, and the modified Jeans outflow velocity at the exobase is only $v_{\text{J} \infty} =$~1.9~$\times$~10$^{-6}$ m~s$^{-1}$.  The critical escape parameter in this case is $X_{\infty} =$~26.  In addition to conduction, cooling from H$_3^+$ plays a substantial role in the lower thermosphere.  This can be seen in the shape of the T-P profile (Figure~\ref{fig:tp}) near the 10 nbar level and below where infrared cooling is comparable to the stellar heating rate. 

At 0.4 AU the exospheric temperature is 2840 K i.e., only slightly higher than at 0.5 AU, and the critical escape parameter is $X_{\infty} =$~20.  Inward from 0.4 AU, the atmosphere undergoes an interesting transition.  For instance, at 0.2 AU (Figure~\ref{fig:tp}) the maximum temperature is much higher than at 0.4 AU, reaching about 10,300 K at 1.4 R$_{\text{p}}$, and formally the exobase is located above the 16 R$_{\text{p}}$ upper boundary of the model.  Above 1.4 R$_{\text{p}}$ the temperature also decreases with altitude.  This is because heating at high altitudes is no longer balanced by conduction -- instead it is balanced by `adiabatic' cooling that is associated with the expansion and escape of the atmosphere.  In this sense the model at 0.2 AU is qualitatively similar to HD209458b where the same behavior has been predicted by several previous models \citep[e.g.,][]{yelle04,garciamunoz07,koskinen12a,koskinen12b}.  Based on these changes in the location of the exobase and the energy balance, we argue that the transition to `hydrodynamic' escape occurs near 0.3 AU.  We note that a similar transition was identified before by \citet{tian08a,tian08b} in the context of the early terrestrial atmosphere.  


\subsection{Escape rates and mixing}
\label{subsc:escape}     

The energy-limited loss rate in the context of extrasolar planets is often formulated as \citep{erkaev07,lammer09}:
\begin{equation}
\dot{L} = \frac{\eta \pi r_{\text{EUV}}^2 F_{\text{EUV}}}{\Phi_0}
\label{eq:energy_limit}
\end{equation}
where $r_{\text{EUV}}$ is the radius of the EUV heating peak, $F_{\text{EUV}}$ is the stellar EUV flux, $\Phi_0$ is the gravitational potential, and $\eta$ is referred to as the `heating efficiency'.  We believe that the use of `heating efficiency' here can be misleading\footnote{Note that heating efficiency is often understood as the fraction of solar energy that heats the atmosphere.}.  For instance, the peak midday heating efficiency of the Earth's thermosphere is 50--55 \% \citep{torr81} but only a tiny fraction of the energy that heats the thermosphere powers escape.  Instead, the heating is mostly balanced by downward heat conduction, radiative cooling and to some degree by circulation.  Under these circumstances a reasonable estimate of the temperature profile can be derived analytically from a simple balance between stellar heating and conduction \citep[e.g.,][]{gross72,lammer03}.  Naturally these considerations are different on Earth in any case because the escape of lighter species such as H and H$_2$ is diffusion-limited, but the discussion here illustrates the general limitations of the energy limit.

With a proper choice of $\eta$, $r_{\text{EUV}}$ and $\Phi_0$, equation~(\ref{eq:energy_limit}) is always accurate and with $\eta =$~1 it yields an upper limit on thermal escape rates.  It is thus be better to call $\eta$ `mass loss efficiency' rather than `heating efficiency'.  This better reflects the fact that equation~(\ref{eq:energy_limit}) describes a balance between external heating and cooling by adiabatic expansion \citep[e.g.,][]{yelle04}.  It also removes any apparent disagreement between equation~(\ref{eq:energy_limit}) and our results at 0.4--1 AU in Section~\ref{subsc:tempvel}.  If one accounts for the fact that heating is mostly balanced by downward conduction, the mass loss efficiency is much lower than the heating efficiency.  Despite the consistency introduced by the new definition of $\eta$, equation~(\ref{eq:energy_limit}) is actually not very useful unless adiabatic cooling really is important.  It may always be possible to tune the mass loss efficiency to force the equation to agree with modeled or observed mass loss rates, but this does not capture the relevant physics in all cases.  Thus the escape mechanism must also be studied in detail before energy-limited escape can be assumed.      

The globally averaged mass loss rates are 3.4~$\times$~10$^{-6}$ kg~s$^{-1}$ and 6.3~$\times$~10$^{-5}$ kg~s$^{-1}$ at 0.5 AU and 0.4 AU, respectively.  Needless to say, these mass loss rates are irrelevant to the long term evolution of the atmosphere.  At 0.2 AU the mass loss rate is 8~$\times$~10$^{5}$ kg~s$^{-1}$.  In this case the heating efficiency is comparable, although not identical, to the mass loss efficiency.  According to our simulations, the heating efficiency is about 8.5 \% at 0.2 AU.  This relatively low value arises because cooling from H$_3^+$ is important in the lower thermosphere below the H$_2$/H transition.  The T-P profile at 0.1 AU is qualitatively similar to the T-P profile at 0.2 AU, and the maximum temperature is about 10,900 K.  The mass loss rate at 0.1 AU is about 6~$\times$~10$^{6}$ kg~s$^{-1}$.  The heating efficiency at 0.1 AU is only 22 \% because H$_3^+$ still cools the lower thermosphere.  As we argue below, this cooling effect becomes negligible within 0.1 AU and thus the heating efficiency for HD209458b increases to 44 \%.  As before, we obtained a mass loss rate of 4.1~$\times$~10$^7$ kg~s$^{-1}$ from our reference model of HD209458b \citep{koskinen12a,koskinen12b}.  

Our results show that heavy species such as C, O, and Si that collide frequently with H and H$^+$ escape the atmosphere of HD209458b with nearly uniform mixing ratios in the thermosphere \citep{koskinen12a,koskinen12b}.  There is an obvious interest in estimating the degree to which this is true on other EGPs because escape can affect the composition and thus the evolution of the atmosphere.  For example, the thermal mass loss rate predicted by us and previous models of HD209458b \citep[e.g.,][]{yelle04} implies that the planet has lost less than 1 \% of its mass during the lifetime of the stellar system.  However, it also implies that HD209458b loses the equivalent mass of its whole atmosphere above the 1 bar level approximately every 800,000 years. 

The crossover mass equation given by \citet{hunten87} can be used to derive a rough estimate of the limiting mass loss rate $\dot{M}_{\text{lim}}$ that is required for a species with mass $M_c$ (in units of $m_\text{H}$) to escape with H due to neutral-neutral collisions:
\begin{equation}
\dot{M}_{\text{lim}} = 4 \pi m_{\text{H}}^2 G M_{\text{p}} (M_c - 1) \frac{n D_c}{k T}
\label{eq:mloss}
\end{equation}
where $D_c$ is the mutual diffusion coefficient for species $c$ with H.  At a temperature of 7,200 K \citep{koskinen12a,koskinen12b}, the mass loss rate required to mix He is about 10$^6$ kg~s$^{-1}$, while the mass loss rate required to mix C and O is 4--6~$\times$~10$^6$ kg~s$^{-1}$.  Given that we estimate a mass loss rate of 6~$\times$~10$^{6}$ kg~s$^{-1}$ at 0.1 AU, progressively heavier neutral species escape the atmosphere with decreasing orbital distance from 0.1 AU.  Because carbon species are effective in removing H$_3^+$, this is another indication that infrared cooling is not effective in the lower thermospheres of close-in EGPs.  Instead H$_2$ is dissociated near the 1 $\mu$bar level and the temperature increases rapidly with altitude above this level.                                          

\subsection{The effect of gravitational potential}
\label{subsc:gravity}  

The escape mechanism depends on the gravitational potential $\Phi$ through its dependence on $X$.  The energy-limited escape rate given by equation~(\ref{eq:energy_limit}) and the crossover mass loss rate given by equation~(\ref{eq:mloss}) also depend on the gravitational potential.  For example, \citet{koskinen09} pointed out that the atmosphere of heavy planets such as HD17156b is not likely to undergo rapid escape even at close-in orbits.  We explored the effect of $\Phi$ in more detail by generating models at the same orbital distances as before but with different values of the surface potential ranging from 0.1 $\Phi_{\text{J}}$ to 3 $\Phi_{\text{J}}$, where $\Phi_{\text{J}}$ is the surface potential of Jupiter.  In order to illustrate the results, we discuss them in the context of a few well known transiting EGPs.  For instance, GJ3470b is a nearby Neptune-size planet orbiting an M dwarf that has a gravitational potential of about 0.1 $\Phi_{\text{J}}$.  The gravitational potential of HD209458b is 0.5 $\Phi_{\text{J}}$ i.e., similar to Saturn, whereas the gravitational potential of HD189733b is identical to Jupiter.  The gravitational potential of HD17156b, on the other hand, is 2.9 $\Phi_{\text{J}}$.  More details on these planets are available, for instance, at www.exoplanet.eu.

The model for HD189733b at 0.2 AU is qualitatively similar to the corresponding model for HD209458b, with a peak temperature of 10,100 K.  However, at 0.3 AU the model for HD189733b is similar to the model of HD209458b at 0.5 AU, with an isothermal temperature of only 2,700 K near the exobase.  This means that with $\Phi = \Phi_{\text{J}}$ the atmosphere enters the rapid escape regime between 0.2 AU and 0.3 AU i.e., slightly inward from the corresponding transition for a planet with $\Phi =$~0.5 $\Phi_{\text{J}}$.  We note that the previous calculations of \citet{koskinen07b}, that also used $\Phi = \Phi_{\text{J}}$, placed this transition between 0.1 AU and 0.2 AU.  The difference between the calculations here and the previous results could be due to circulation.  With a slow rotation rate approaching tidal synchronization that was assumed by \citet{koskinen07b}, the GCM at 0.2 AU develops strong day-night circulation at high altitudes that leads to upwelling in the dayside.  This upwelling replenishes H$_2$ and helps to maintain the stability of the atmosphere.  However, a faster rotation rate with a period of 24 hours or less disturbs this type of circulation and leads to a transition to the rapid escape regime between 0.2 AU and 0.3 AU \citep{koskinen08}, in agreement with the 1D globally averaged simulations presented here.  It should be noted that while close-in EGPs are assumed to be rotationally synchronized with their orbital periods \citep[e.g.][]{guillot96}, there is no reason to assume that this is the case farther out between 0.2 and 0.3 AU. 

With the surface potential increased to $\Phi =$~3 $\Phi_{\text{J}}$, the atmosphere at 0.2 AU does not undergo hydrodynamic escape.  Instead, the exospheric temperature is only 2,820 K with an exobase at 1.08 R$_{\text{p}}$ and $X_{\infty} =$~207.  At 0.1 AU, H$_2$ dissociates rapidly above the 1 $\mu$bar level and the exospheric temperature increases to 12,400 K.  However, even in this case the exobase is at 1.5 R$_{\text{p}}$ with $X_{\infty} =$~33 and the atmosphere does not escape hydrodynamically.  At 0.05 AU, the exospheric temperature is 14,340 K and $X_{\infty} =$~25.  These results confirm the finding of \citet{koskinen09} that heavy planets such as HD17156b do not undergo rapid escape even at close-in orbits, despite the fact that H$_2$ is dissociated in their upper atmospheres.  

Escape is less effective on heavier planets partly because radiative cooling offsets the stellar XUV heating near the heating peak.  In our new calculations we include cooling due to recombination but H Lyman $\alpha$ cooling can also contribute \citep{murrayclay09}, keeping the temperature in the thermosphere close to 10,000 K.  These cooling effects play a more significant role in the energy balance on planets with higher gravity because the temperature in the thermosphere on such planets is not high enough for adiabatic cooling to become important.  Further research is required to study the role of these cooling mechanisms in detail though as they are rarely included in models of giant planet atmospheres.  In any case, our results show that a high speed wind containing potassium atoms from the exosphere of another similar planet HD80606b, as proposed by \citet{colon12}, is unlikely to be realistic.  The same considerations do not apply to planets such as GJ3470b with $\Phi =$~0.1 $\Phi_{\text{J}}$.  We found that such planets undergo a transition to rapid escape much farther out, and the exobase extends to very high altitude even at 1 AU. 

Naturally, surface gravity also affects $\dot{M}_{\text{lim}}$, which is directly proportional to the mass of the planet.  Thus the mass loss rate required to mix, say, C into the thermosphere of HD189733b is 1.7 times higher than on HD209458b.  On HD17156b the corresponding rate is 4.6 times higher than on HD209458b and on GJ3470 it is 16 times lower than on HD209458b.  Thus the collisional mixing of heavy species is much more likely on planets with a low surface gravity than on heavy planets such as HD17156b.  Given that the escape rates on HD17156b or HD80606b are likely to be comparable to the Jeans escape rate even at the periastron near 0.05 AU, we do not expect that substantial abundances of heavy elements escape from their atmospheres.  Instead, heavy species on these planets are likely to be diffusively separated below the exobase.                               

\section{Discussion and Conclusions}
\label{sc:discussion}    

The study of extrasolar planet atmospheres is still an emerging field.  This means that there are many uncertainties in both the models and even the interpretation of the observations.  For instance, we used T-P profiles in the lower atmosphere that are based on models that make many simplifying assumptions about the composition and radiative transfer \citep{sudarsky03,burrows04}.  Although these models account for condensation, cloud formation and the scattering of radiation, they ignore the effects of photochemistry and non-LTE radiative transfer that can, for instance, produce strong emission features on close-in planets \citep{swain10,waldmann12}.  The temperature profile in the thermosphere, on the other hand, depends on the photoelectron heating efficiencies and the H$_3^+$ infrared cooling rates.  Fully self-consistent calculations of the heating efficiency do not yet exist, and H$_3^+$ emissions have not been detected on EGPs.  In fact, our estimates indicate that these emissions are undetectable with current instruments despite the fact that they can play a substantial role in the energy balance of EGPs.   

The composition and T-P profiles are also affected by dynamics and turbulent mixing.  There are no observational constraints on the impact of dynamics on the composition, and only a few constraints on circulation in general.  Yet dynamics can play a large role in the atmosphere, controlling processes such as cloud formation and the mixing ratios of heavy elements in the upper atmosphere.  Photochemistry, dynamics, and thermal structure are all driven by the stellar flux.  It is surprising how little is known about the properties of the host stars and the details of their spectra.  The activity cycle of even other Sun-like stars is not well understood although it is likely to affect both the atmospheres and the interpretation of transit observations.  Even less is known about stars of other spectral type.  For instance, M stars exhibit a range of activity levels in general, and can be highly variable even on short time scales \citep{france12}.  

Despite the uncertainties, we are able to identify some robust qualitative results.  First, there are two distinct regimes of atmospheric escape that are separated by a relatively sharp transition controlled by the stellar flux.  In the `stable' regime, stellar heating is balanced by downward conduction and/or radiative cooling.  In this case the mass loss rate is typically many orders of magnitude smaller than the energy-limited escape rate that is based on the thermospheric heating efficiency.  In the `unstable' regime stellar heating is efficient enough to cause a significant expansion of the atmosphere.  As a result, stellar heating is balanced by adiabatic cooling and the temperature decreases with altitude above the EUV heating peak.  In this regime the mass loss rate is comparable to the energy-limited mass loss rate that is based on the thermospheric heating efficiency.  The transition between these two regimes on close-in EGPs is sharp because it is in many cases driven by the dissociation of H$_2$.  The H$_2$/H dissociation front is sharp both as a function of altitude in the atmospheric models and in terms of orbital distance. 

In addition to stellar flux and composition, the escape mechanism depends on the surface gravitational potential.  We find that planets such as HD209458b with $\Phi =$~0.5 $\Phi_{\text{J}}$ are stable at orbital distances greater than about 0.3 AU.  In our calculations this corresponds to an integrated EUV flux of $F_{\text{lim}} =$~44 mW~m$^{-2}$ at wavelengths shorter than 912 \AA.  For planets like Jupiter this limit is closer at 0.2 AU ($F_{\text{lim}} =$~0.1 W~m$^{-2}$).  The atmospheres of heavier planets such as HD17156b are stable even at 0.05 AU ($F_{\text{lim}} =$~1.6 W~m$^{-2}$), and unstable only near 0.01 AU.  The atmospheres of planets such as GJ3470b are much more likely to escape hydrodynamically, even near 1 AU.  The escape of heavy species is controlled by the escape rate of hydrogen.  For a planet such as HD209458b, the energy-limited escape rate is high enough to cause species more massive than C to escape within 0.1 AU.  The escape of heavy species from planets such as HD17156b is unlikely whereas planets such as GJ3470b can start losing heavy species within 1 AU.  We note here that we did not calculate photochemistry in the lower atmosphere for planets with different $\Phi$, assuming that the changes in the mass and radius of the planet do not affect the H$_2$/H transition significantly.     

\begin{figure}
  \epsscale{0.8}
  \plotone{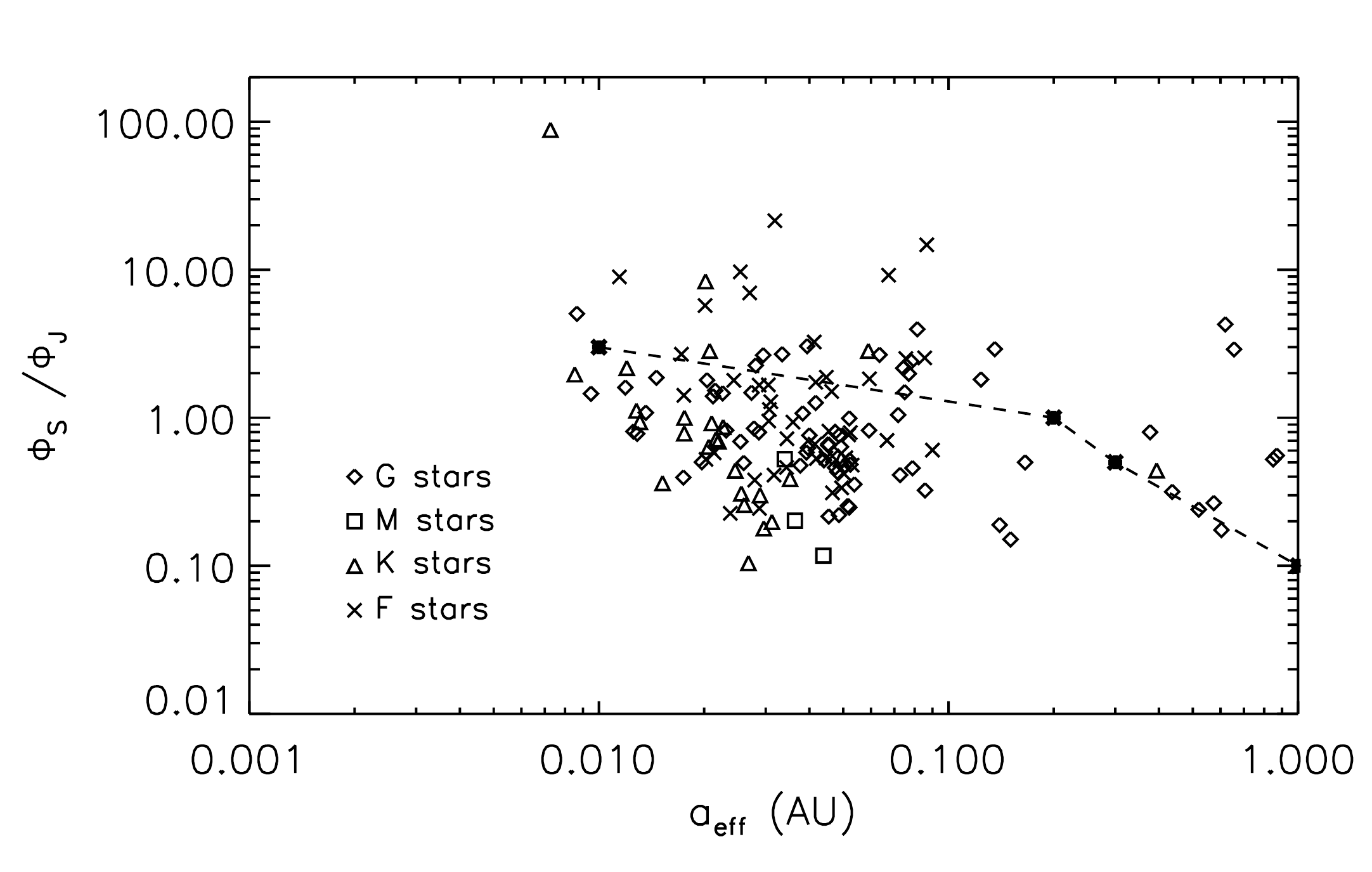}
  \caption{Surface potential of EGPs with known radius, mass, and the spectral type of the host star as a function of effective orbital distance (see text).  The dashed line shows the transition between unstable escape-dominated (to the left) and stable (to the right) regimes as defined in the text.  The stars show the values of $\Phi$ for which simulations were performed.}
  \label{fig:egps}
\end{figure}

To illustrate the results, Figure~\ref{fig:egps} shows the gravitational potential $\Phi = G M_{\text{p}}/R_{\text{p}}$ of currently known transiting EGPs (planets with a mass higher than 10 M$_{\text{E}}$) as a function of effective orbital distance.  This is the orbital distance in the Solar System where the planets would receive the same EUV flux as they do currently, given an estimate of the EUV flux of their host stars.  For G stars, we used the solar flux (4 mW~m$^{-2}$ at 1 AU).  When an estimate of the stellar age exists, we scaled the flux according to equation (1) of \citet{ribas05}.  For M, K, and F stars we used the simple scaling proposed by \citet{lecavelier07} without accounting for a possible age dependence.  We excluded planets for which the spectral type of the host star is not listed at www.exoplanet.eu.  The names of the planets in Figure~\ref{fig:egps} can be made available on request.  

Our results indicate that most of the transiting EGPs have rapidly escaping atmospheres, and heavy species are expected to escape with H on many of the currently known planets.  However, this is not true of the heavier planets such as HD17156b or HD80606b, even if they come close to the host star.  There is also an intermediate region in Figure~\ref{fig:egps} where hydrogen is escaping rapidly but heavier species are still likely to be diffusively separated and not escape the atmosphere with large abundances.  It is interesting that there are no planets within 0.01--0.03 AU.  \citet{garciamunoz07} pointed out that the mass loss rate increases very steeply with decreasing orbital distance within 0.015 AU due to stellar gravity.  At such orbital distance the interaction of the planet with the stellar wind is also expected to be more violent.  In this context it is important to note that our work does not include stellar gravity or any additional escape mechanisms such as erosion by the stellar wind.  Our results are intended to provide constraints on the thermal escape rate and mechanism that can be used as a base of more complex models in the future when more detailed observations are available to characterize the complex physics of atmospheric escape. 

Based on estimates of the stellar flux and $\Phi$, the limits given in Figure~\ref{fig:egps} can be used to determine the thermal escape mechanism for different EGPs.  We note, however, that the limits for the different escape regimes here are based on models with a relatively few values of $a_{\text{eff}}$ and $\Phi$, and we recommend that interested readers contact the authors for separate models of specific targets if more accurate results are required.  We also stress that only a detailed comparison of the simulations with actual observations can validate the models, and future efforts in characterizing the upper atmospheres of EGPs should concentrate on obtaining multiple observations of both the host stars and the planets in different systems, and designing new techniques and instruments to make these observations possible.                                         

\acknowledgments     

The calculations in this paper relied on the High Performance Astrophysics Simulator (HiPAS) at the University of Arizona, and the University College London Legion High Performance Computing Facility, which is part of the DiRAC Facility jointly funded by STFC and the Large Facilities Capital Fund of BIS.  SOLAR2000 Professional Grade V2.28 irradiances are provided by Space Environment Technologies.  This research was supported by the National Science Foundation (NSF) grant AST1211514. 



\begin{thebibliography}{}
\bibitem[Adams(2011)]{adams11} Adams, F. C., 2011.  Magnetically controlled outflows from hot Jupiters.  Astrophys. J., 730, 27.
\bibitem[Burrows et al.(2004)]{burrows04} Burrows, A., Sudarsky, D., Hubeny, I., 2004.  Spectra and diagnostics for the direct detection of wide-separation extrasolar giant planets.  Astrophys. J., 609, 407--416.
\bibitem[Colon et al.(2012)]{colon12} Col\'on, K. D., et al., 2012.  Probing potassium in the atmosphere of HD80606b with tunable filter transit spectrophotometry from the Gran Telescopio Canarias.  Mon. Not. R. Astron. Soc., 419, 2233--2250. 
\bibitem[Erkaev et al.(2007)]{erkaev07} Erkaev, N. V., et al., 2007.  Roche lobe effects on the atmospheric loss from "Hot Jupiters".  Astron. Astrophys., 472, 329--334.
\bibitem[France et al.(2012)]{france12} France, K., et al., 2012.  The ultraviolet radiation environment around M dwarf exoplanet host stars.  arXiv:1212.4833v1. 
\bibitem[Fossati et al.(2010)]{fossati10} Fossati, L., et al., 2010.  Metals in the exosphere of the highly irradiated planet WASP-12b.  Astrophys. J. Lett., 714, L222--L227.
\bibitem[Garcia Munoz(2007)]{garciamunoz07} Garcia Munoz, A.  2007.  Physical and chemical aeronomy of HD209458b.  Plan. Sp. Sci., 55, 1426--1455.
\bibitem[Gross(1972)]{gross72} Gross, S. H., 1972.  On the exospheric temperature of hydrogen-dominated planetary atmospheres.  J. Atmos. Phys., 29, 214--218.  
\bibitem[Guillot et al.(1996)]{guillot96} Guillot, T., Burrows, A., Hubbard, W. B., Lunine, J. I., Saumon, D., 1996.  Giant planets at small orbital distances.  Astrophys. J. Lett., L35--L38. 
\bibitem[Hunten et al.(1987)]{hunten87} Hunten, D. M., Pepin, R. O., Walker, J. C. G., 1987.  Mass fractionation in hydrodynamic escape.  Icarus, 69, 532--549.
\bibitem[Koskinen et al.(2007a)]{koskinen07a} Koskinen, T. T., Aylward, A. D., Miller, S., Smith, C. G. A., 2007a.  A thermospheric circulation model for extrasolar giant planets.  Astrophys. J., 661, 515--526.
\bibitem[Koskinen et al.(2007b)]{koskinen07b} Koskinen, T. T., Aylward, A. D., Miller, S., 2007b.  A stability limit for the atmospheres of giant extrasolar planets.  Nature, 450, 845--848.
\bibitem[Koskinen et al.(2008)]{koskinen08} Koskinen, T. T., 2008.  Stability of short period exoplanets.  PhD thesis.  University College London, London, England.  
\bibitem[Koskinen et al.(2009)]{koskinen09} Koskinen, T. T., Aylward, A. D., Miller, S., 2009.  The upper atmosphere of HD17156b.  Astrophys. J., 693, 868--885.   
\bibitem[Koskinen et al.(2010)]{koskinen10} Koskinen, T. T., Yelle, R. V., Lavvas, P., Lewis, N., 2010a.  Characterizing the thermosphere of HD209458b with UV transit observations.  Astrophys. J., 723, 116--128.
\bibitem[Koskinen et al.(2010b)]{koskinen10b} Koskinen, T. T., Cho, J. Y-K., Achilleos, N., Aylward, A. D., 2010b.  Ionization of extrasolar giant planet atmospheres.  Astrophys. J., 722, 178--187.
\bibitem[Koskinen et al.(2013a)]{koskinen12a} Koskinen, T. T., Harris, M. J., Yelle, R. V., Lavvas, P., 2013a.  The escape of heavy atoms from the ionosphere of HD209458b. I. A photochemical-dynamical model of the thermosphere.  Icarus, 226, 1678--1694.
\bibitem[Koskinen et al.(2013b)]{koskinen12b} Koskinen, T. T., Yelle, R. V., Harris, M. J., Lavvas, P., 2013b.  The escape of heavy atoms from the ionosphere of HD209458b. II. Interpretation of the observations.  Icarus, 226, 1695--1708.
\bibitem[Koskinen et al.(2013c)]{koskinen13} Koskinen, T. T., et al., 2013c.  The density and temperature structure near the exobase of Saturn from Cassini UVIS solar occultations.  Icarus, 226, 1318--1330.
\bibitem[Lammer et al.(2003)]{lammer03} Lammer, H., et al., 2003.  Atmospheric loss of exoplanets resulting from stellar X-ray and extreme-ultraviolet heating.  Astrophys. J. Lett., 598, L121--L124. 
\bibitem[Lammer et al.(2009)]{lammer09} Lammer, H., et al., 2009.  Determining the mass loss limit for close-in exoplanets: What can we learn from transit observations?  Astron. Astrophys., 506, 399--410. 
\bibitem[Lavvas et al.(2008a)]{lavvas08a} Lavvas, P., Coustenis, A., Vardavas, I. M., 2008a.  Coupling photochemistry with haze formation in Titan's atmosphere, part I: Model description.  Plan. Space Sci., 56, 27--66.
\bibitem[Lavvas et al.(2008b)]{lavvas08b} Lavvas, P., Coustenis, A., Vardavas, I. M., 2008b.  Coupling photochemistry with haze formation in Titan's atmosphere, part II: Results and validation with Cassini/Huygens data.  Plan. Space Sci., 56, 67--99.
\bibitem[Lecavelier des Etangs(2007)]{lecavelier07} Lecavelier des Etangs, A., 2007.  A diagram to determine the evaporation status of extrasolar planets.  Astron. Astrophys., 461, 1185--1193.   
\bibitem[Lecavelier des Etangs et al.(2010)]{lecavelier10} Lecavelier des Etangs, A., et al., 2010.  Evaporation of the planet HD189733b observed in H I Lyman $\alpha$.  Astron. Astrophys., 514, A72.
\bibitem[Lecavelier des Etangs et al.(2012)]{lecavelier12} Lecavelier des Etangs, A., et al., 2012.  Temporal variations in the evaporating atmosphere of the exoplanet HD189733b.  Astron. Astrophys. Lett., 543, L4.
\bibitem[Linsky et al.(2010)]{linsky10} Linsky, J. L., et al.  2010.  Observations of mass loss from the transiting exoplanet HD209458b.  Astrophys. J., 717, 1291--1299.
\bibitem[Miller et al.(2005)]{miller05} Miller, S., Aylward, A. D, Millward, G., 2005.  Giant planet ionospheres and thermospheres: The importance of ion-neutral coupling.  Space Sci. Rev., 116, 319--343. 
\bibitem[Moses et al.(2011)]{moses11} Moses, J. I., et al., 2011.  Disequilibrium carbon, oxygen, and nitrogen chemistry in the atmospheres of HD189733b and HD209458b.  Astrophys. J., 737, 15.
\bibitem[M\"uller-Wodarg et al.(2006)]{mullerwodarg06} M\"uller-Wodarg, I. C. F., Mendillo, M., Yelle, R. V., Aylward, A. D., 2006.  A global circulation model of Saturn's thermosphere.  Icarus, 180, 147--160.
\bibitem[Murray-Clay et al.(2009)]{murrayclay09} Murray-Clay, R., Chiang, E. I., Murray, N., 2009.  Atmospheric escape from hot Jupiters.  Astrophys. J., 693, 23--42.
\bibitem[Parker(1958)]{parker58} Parker, E. N., 1958.  Dynamics of the interplanetary gas and magnetic fields.  Astrophys. J., 128, 664--676. 
\bibitem[Ribas et al.(2005)]{ribas05} Ribas, I., Guinan, E. F., G\"udel, M., Audard, M., 2005.  Evolution of the solar activity over time and effects on planetary atmospheres. I. High-energy irradiances (1--1700 \AA).  Astrophys. J., 622, 680--694.
\bibitem[Seiff et al.(1998)]{seiff98} Seiff, A., et al., 1998.  Thermal structure of Jupiter's atmosphere near the edge of a 5 $\mu$m hot spot in the north equatorial belt.  J. Geophys. Res., 103, 22857--22889.
\bibitem[Showman et al.(2009)]{showman09} Showman, A., et al., 2009.  Atmospheric circulation of hot Jupiters: Coupled radiative-dynamical general circulation model simulations of HD189733b and HD209458b.  Astrophys. J., 699, 564--584.
\bibitem[Smith et al.(2007)]{smith07} Smith, C. G. A., Aylward, A. D., Millward, G. H., Miller, S., Moore, L. E., 2007.  An unexpected cooling effect in Saturn's upper atmosphere.  Nature, 445, 399--401.
\bibitem[Stone and Proga(2009)]{stone09} Stone, J. M., Proga, D., 2009.  Anisotropic winds from close-in extrasolar planets.  Astrophys. J., 694, 205--213.  
\bibitem[Strobel(2002)]{strobel02} Strobel, D. F., 2002.  Aeronomic systems on planets, moons, and comets.  In Atmospheres in the solar system: Comparative aeronomy.  Geophys. Monogr. Ser., 130, ed. by M. Mendillo, A. Nagy, H. Waite, 7--22.  AGU, Washington, D. C.
\bibitem[Sudarsky et al.(2003)]{sudarsky03} Sudarsky, D., Burrows, A., Hubeny, I., 2003.  Theoretical spectra and atmospheres of extrasolar giant planets.  Astrophys. J., 588, 1121--1148.
\bibitem[Swain et al.(2010)]{swain10} Swain, M. R., et al., 2010.  A ground-based near-infrared emission spectrum of the exoplanet HD189733b.  Nature, 463, 637--639.
\bibitem[Tian et al.(2008a)]{tian08a} Tian, F., Kasting, J. F., Liu, H-L., Roble, R. G., 2008a.  Hydrodynamic planetary thermosphere model: 1. Response of the Earth's thermosphere to extreme solar EUV conditions and the significance of adiabatic cooling.  J. Geophys. Res., 113, E05008.
\bibitem[Tian et al.(2008b)]{tian08b} Tian, F., Solomon, S. C., Qian, L., Lei, J., Roble, R. G., 2008b.  Hydrodynamic planetary thermosphere model: 2. Coupling of an electron transport/energy deposition model.  J. Geophys. Res., 113, E07005.
\bibitem[Trammell et al.(2011)]{trammell11} Trammell, G. B., Arras, P., Li, Z.-Y., 2011.  Hot Jupiter magnetospheres.  Astrophys. J., 728, 152.
\bibitem[Torr et al.(1981)]{torr81} Torr, M. R., Richards, P. G., Torr, D. G., 1981.  Solar EUV energy budget of the thermosphere.  Adv. Space Res., 1, 53--61.
\bibitem[Vidal-Madjar et al.(2003)]{vidalmadjar03} Vidal-Madjar, A., et al.,  2003.  An extended upper atmosphere around the extrasolar giant planet HD209458b.  Nature, 422, 143--146.
\bibitem[Vidal-Madjar et al.(2004)]{vidalmadjar04} Vidal-Madjar, A., et al.,  2004.  Detection of oxygen and carbon in the hydrodynamically escaping atmosphere of the extrasolar planet HD209458b.  Astrophys. J. Lett., 604, L69--L72.
\bibitem[Volkov et al.(2011a)]{volkov11a} Volkov, A. N., Johnson, R. E., Tucker, O. J., Erwin, J. T., 2011a.  Thermally driven atmospheric escape: Transition from hydrodynamic escape to Jeans escape.  Astrophys. J. Lett., 729, L24.
\bibitem[Volkov et al.(2011b)]{volkov11b} Volkov, A. N., Tucker, O. J., Erwin, J. T., Johnson, R. E., 2011b.  Kinetic simulations of thermal escape from a single component atmosphere.  Phys. Fluids, 23, 066601.
\bibitem[Waldmann et al.(2012)]{waldmann12} Waldmann, I. P., et al., 2012.  Ground-based near-infrared emission spectroscopy of HD189733b.  Astrophys. J., 744, 35. 
\bibitem[Woods and Rottman(2002)]{woods02} Woods, T. N., Rottman, G. J., 2002.  Solar ultraviolet variability over time periods of aeronomical interest.  In: Mendillo, M., et al. (Eds.), Comparative aeronomy in the solar system.  American Geophysical Union Monograph.
\bibitem[Yelle(2004)]{yelle04} Yelle, R. V., 2004.  Aeronomy of extra-solar giant planets at small orbital distances.  Icarus, 170, 167--179.
\bibitem[Yelle(2006)]{yelle06} Yelle, R. V., 2006.  Corrigendum to "Aeronomy of extra-solar giant planets at small orbital distances".  Icarus, 183, 508.
\end{thebibliography}
\end{document}